\documentclass[5p]{elsarticle}

\usepackage{amssymb}
\usepackage{amsmath}
\usepackage{graphics}
\usepackage{float}
\usepackage{flushend}

\begin{document}

\begin{frontmatter}

\title{Critical Point Estimation and Long-Range Behavior in the One-Dimensional XY Model Using Thermal Quantum and Total Correlations}

\author[]{B. \c{C}akmak\corref{cor1}}
\ead{cakmakb@sabanciuniv.edu}
\cortext[cor1]{Corresponding author : Tel: +90 2164839000 / 2325}

\author{G. Karpat}

\author{Z. Gedik}

\address{Faculty of Engineering and Natural Sciences, Sabanci University, Tuzla, Istanbul 34956, Turkey}

\begin{abstract}
We investigate the thermal quantum and total correlations in the anisotropic XY spin chain in transverse field. While we adopt concurrence and geometric quantum discord to measure quantum correlations, we use measurement-induced nonlocality and an alternative quantity defined in terms of Wigner-Yanase information to quantify total correlations. We show that the ability of these measures to estimate the critical point at finite temperature strongly depend on the anisotropy parameter of the Hamiltonian. We also identify a correlation measure which detects the factorized ground state in this model. Furthermore, we study the effect of temperature on long-range correlations.
\end{abstract}


\end{frontmatter}

\section{Introduction}

The notion of entanglement, on top of being considered as the resource of quantum computation, quantum cryptography and quantum information processing, has also proved to be very useful in analyzing the behavior of various condensed matter systems [1, 2]. However, entanglement is not the only kind of meaningful correlation present in quantum systems. Quantum discord [3, 4], defined as the discrepancy between quantum versions of two classically equivalent expressions for mutual information, has been demonstrated to be a novel resource for quantum computation [5-7]. Following the path paved by the authors of Ref. [3, 4], several new quantifiers of quantum correlations and non-locality, that are more general than entanglement, have been proposed recently [8-11].

Quantum phase transitions (QPTs) are sudden changes occurring in the ground states of many-body systems when one or more of the physical parameters of the system are continuously varied at absolute zero temperature [12]. Th-\\ese radical changes, which strongly affect the macroscopic properties of the system, are manifestations of quantum fluctuations. Despite the fact that reaching absolute zero temperature is practically impossible, QPTs might still be observed at sufficiently low temperatures, where thermal fluctuations are not significant enough to excite the system from its ground state. In recent years, the methods of quantum information theory have been widely applied to quantum critical systems. In particular, entanglement and quantum discord (QD) have been shown to identify the critical points (CPs) of QPTs with success in several different critical spin chains, both at zero [13-25] and finite temperature [26-28]. It has also been noted that unlike pairwise entanglement, which is typically short ranged, QD does not vanish even for distant spin pairs [18].

An interesting aspect of quantum spin chains in transverse magnetic field is the occurrence of a non-trivial factorized ground state [29]. In order to gain a complete understanding of these factorized states, the effects of spontaneous symmetry breaking (SSB) should be considered [30-32]. In fact, concurrence is known to signal the factorization point of the anisotropic XY chain corresponding to a product ground state [32]. Moreover, it has been demonstrated that QD is also able to detect such points, provided that either SSB is taken into account or QD is calculated for different spin distances [33,34]. In the latter case, the factorization point appears via a single intersection of the curves of QD.

In this work, we investigate the pairwise thermal quantum and total correlations in the one-dimensional anisotro-\\pic spin-$1/2$ XY chain in transverse magnetic field. As a measure of genuine quantum correlations, we utilize entanglement quantified by concurrence [35,36], and a very recently proposed observable measure (OMQC) [11], which is a simplified version of geometric measure of quantum discord (GMQD) [8]. This observable measure has the advantage that it does not require a full tomography of the system, making it experimentally very accessible. On the other hand, in order to quantify non-locality or total correlations in a quantum system, we use measurement-induced nonlocality (MIN) [9], and an alternative new measure defined in terms of Wigner-Yanase skew information (WYSIM) [10]. By comparatively studying the thermal quantum and total correlations in the parameter space of the Hamiltonian for both first and second nearest neighbor spins, we observe that all of these measures are capable of indicating the CP of QPT at absolute zero. When the temperature is slightly above the absolute zero, in the experimentally accessible region, we analyze the ability of correlation measures to correctly estimate the CP of the transition.
It is also remarkable that, among the new measures considered in this work, only WYSIM is able to identify the factorization point of the XY spin chain even if we disregard the effects of SSB. Lastly, we study the long-range correlations of the system and the effect of temperature on these correlations.

This paper is organized as follows. In Section 2, we review the correlation measures used in our work. In Section 3, we introduce the one-dimensional XY spin chain, and discuss the thermal correlations of the system from different perspectives to estimate the CP of the QPT. Section 4 includes the summary of our results.

\section{Correlation Measures}

In this section, we review several measures of total and quantum correlations to be used in our investigation of the one-dimensional XY spin chain. We will commence by introducing MIN which encapsulates more general kind of correlations than quantum non-locality connected with the violation of Bell inequalities [9]. It is defined by (taking into account the normalization)
\begin{equation}
N(\rho^{ab})=2\max_{\Pi^{a}}\|\rho^{ab}-\Pi^{a}(\rho^{ab})\|^{2},
\end{equation}
where the maximum is taken over the von Neumann measurements $\Pi^{a}=\{\Pi_{k}^{a}\}$ that do not change $\rho^{a}$ locally, meaning $\sum_{k}\Pi_{k}^{a}\rho^{a}\Pi_{k}^{a}=\rho^{a}$, and $\|.\|^{2}$ denotes the square of the Hilbert-Schmidt norm. MIN aims to capture the non-local effect of the measurements on the state $\rho^{ab}$ by requiring that the measurements do not disturb the local state $\rho^{a}$. It is always possible to represent a general bipartite state in Bloch basis as
\begin{align}
\rho^{ab} &= \frac{1}{\sqrt{mn}} \frac{I^{a}}{\sqrt{m}} \otimes \frac{I^{b}}{\sqrt{n}}+ \sum_{i=1}^{m^2-1}x_{i}X_{i} \otimes \frac{I^{b}}{\sqrt{n}}  \nonumber \\
     &+\frac{I^{a}}{\sqrt{m}} \otimes \sum_{j=1}^{n^2-1}y_{j}Y_{j} + \sum_{i=1}^{m^2-1}\sum_{j=1}^{n^2-1} t_{ij} X_{i} \otimes Y_{j},
\end{align}
where the matrices $\{X_{i}: i=0,1,\cdots,m^{2}-1\}$ and $\{Y_{j}: j=0,1,\cdots,n^{2}-1\}$, satisfying $\textmd{tr}(X_{k}X_{l})=\textmd{tr}(Y_{k}Y_{l})=\delta_{kl}$, define an orthonormal Hermitian operator basis associated to the subsystems $a$ and $b$, respectively. The components of the local Bloch vectors $\vec{x}=\{x_{i}\}$, $\vec{y}=\{y_{j}\}$ and the correlation matrix $T=t_{ij}$ can be obtained as
\begin{align}
x_{i}  &= \textmd{tr}\rho^{ab}(X_{i} \otimes I^{b})/\sqrt{n}\nonumber, \\
y_{j}  &= \textmd{tr}\rho^{ab}(I^{a} \otimes Y_{j})/\sqrt{m}\nonumber, \\
t_{ij} &=\textmd{tr}\rho^{ab}(X_{i}\otimes Y_{j}).
\end{align}
Although a closed formula for the most general case of bipartite quantum systems is not known, provided that we have a two-qubit system ($m=n=2$), MIN can be analytically evaluated as
\begin{align}
 N(\rho) = \begin{cases}
          2(\textmd{tr}TT^{t}- \frac{1}{\|\vec{x}\|^{2}} \vec{x}^{t}TT^{t}\vec{x}) & \text{if } \vec{x} \neq 0, \\
          2(\textmd{tr}TT^{t}- \lambda_{3}) & \text{if } \vec{x} = 0,
        \end{cases}
\end{align}
where $TT^{t}$ is a $3 \times 3$ dimensional matrix with $\lambda_{3}$ being its minimum eigenvalue, and $\|\vec{x}\|^{2}=\sum_{i}x_{i}^{2}$ with $\vec{x}=(x_{1},x_{2},x_{3})^{t}$. Due to the symmetries of the considered system in this work, the two-spin reduced density matrix is $X$-shaped
\begin{equation}
 \rho^{ab} = \begin{pmatrix}
 \rho_{11} & 0 & 0 & \rho_{14} \\
 0 & \rho_{22} & \rho_{23} & 0 \\
 0 & \rho_{23} & \rho_{22} & 0 \\
 \rho_{14} & 0 & 0 & \rho_{44} \end{pmatrix}.
\end{equation}
Since the local Bloch vector $\vec{x}$ is never zero in our investigation, MIN takes the simple form
\begin{equation}
 N(\rho )=4(\rho_{23}^2+\rho_{14}^2).
\end{equation}

Very recently, a new measure of total correlations has been proposed in Ref. [10] by making use of the notion of Wigner-Yanase skew information
\begin{equation}
I(\rho,X)=-\frac{1}{2}\textmd{tr}[\sqrt{\rho},X]^{2},
\end{equation}
which has been first introduced by Wigner and Yanase [37]. Here $X$ is an observable (an Hermitian operator) and $[.,.]$ denotes commutator. For pure states, $I(\rho,X)$ reduces to the variance $V(\rho,X)=\textmd{tr}\rho X^{2}-(\textmd{tr}\rho X)^2$. Since the skew information $I(\rho,X)$ depends both on the state $\rho$ and the observable $X$, Luo introduced an average quantity in order to get an intrinsic expression
\begin{equation}
Q(\rho)=\sum_{i}I(\rho,X_{i}),
\end{equation}
where $\{X_{i}\}$ is a family of observables which constitutes an orthonormal basis. Global information content of a bipartite quantum system $\rho^{ab}$ with respect to the local observables of the subsystem $a$ can be defined by
\begin{equation}
Q_{a}(\rho^{ab})=\sum_{i}I(\rho^{ab},X_{i}\otimes I^{b}),
\end{equation}
which does not depend on the choice of the orthonormal basis $\{X_{i}\}$. Then, the difference
between the information content of $\rho^{ab}$ and $\rho^{a} \otimes \rho^{b}$ with respect to the local observables of the subsystem $a$ can be adopted as a correlation measure for $\rho^{ab}$,
\begin{align}
F(\rho^{ab}) &= \frac{2}{3}(Q_{a}(\rho^{ab})-Q_{a}(\rho^{a}\otimes\rho^{b})) \nonumber, \\
        &= \frac{2}{3}(Q_{a}(\rho^{ab})-Q_{a}(\rho^{a})),
\end{align}
where we add a normalization factor $2/3$. Despite the fact that the evaluation of most of the measures requires a potentially complex optimization process, $F(\rho^{ab})$ (referred as WYSIM) has the advantage that it can be calculated straightforwardly. At this point, we note that quantum mutual information (QMI) has been widely used as the original measure of total correlations contained in quantum states. Being based on the von Neumann entropy, QMI is a well established measure from the communication perspective, while WYSIM is based on the skew information and has a fundamental role in quantum estimation theory [10]. Although we do not adopt QMI as a measure of total correlations in this work, it has been shown to identify CPs in quantum critical systems [14,38-40].

GMQD has been introduced to overcome the difficulties in the evaluation of the original QD [8]. It measures the nearest distance between a given state and the set of zero-discord states. Mathematically, it is given by
\begin{equation}
D_{G}(\rho^{ab})=2\min_{\chi}\|\rho^{ab}-\chi\|^{2},
\end{equation}
where the minimum is taken over the set of zero-discord states. In a recent work, Girolami et al. have obtained an interesting analytical formula for the GMQD of an arbitrary two-qubit state [11]
\begin{equation}
D_{G}(\rho^{ab})=2(\textmd{tr}S-\max\{k_{i}\}),
\end{equation}
where $S=\vec{x}\vec{x}^{t}+TT^{t}$ and
\begin{equation}
k_{i}=\frac{\textmd{tr}S}{3}+ \frac{\sqrt{6\textmd{tr}S^{2}-2(\textmd{tr}S)^{2}}}{3}\cos\left(\frac{\theta+\alpha_{i}}{3}\right),
\end{equation}
with $\{\alpha_{i}\}=\{0,2\pi,4\pi\}$ and $\theta=\arccos\{(2\textmd{tr}S^{3}-9\textmd{tr}S\textmd{tr}S^{2}+9\textmd{tr}S^{3})\sqrt{2/(3\textmd{tr}S^{2}
-(\textmd{tr}S)^{2})^{3}}\}$. Furthermore, observing that $\cos\left(\frac{\theta+\alpha_{i}}{3}\right)$ reaches its maximum for $\alpha_{i}=0$ and choosing $\theta$ to be zero, they have found a very tight lower bound to the GMQD, given by
\begin{equation}
Q(\rho^{ab})=\frac{2}{3}(2\textmd{tr}S-\sqrt{6\textmd{tr}S^{2}-2(\textmd{tr}S)^{2}}).
\end{equation}
This quantity (referred as OMQC) can be regarded as a meaningful measure of quantum correlations on its own and it has the desirable feature that it needs no optimization procedure. Besides being easier to manage than the original GMQD, it can be measured by performing seven local projections on up to four copies of the state. Thus, $Q(\rho)$ is also very experimentally friendly since one does not need to perform a full tomography of the state.

Lastly, we utilize concurrence to quantify the entanglement content of two-qubit density matrices [35,36]. In order to evaluate concurrence, one first needs to calculate the time-reversed or spin-flipped density matrix $\tilde{\rho}$ which is given by
\begin{equation}
\tilde{\rho}=(\sigma^{y}\otimes\sigma^{y})\rho^{*}(\sigma^{y}\otimes\sigma^{y}).
\end{equation}
Here $\sigma^{y}$ is the Pauli spin operator and $\rho^{*}$ is obtained from $\rho$ via complex conjugation. Then, concurrence reads
\begin{equation}
C(\rho)=\max \left\{ 0,\sqrt{\lambda_{1}}-\sqrt{\lambda_{2}}-\sqrt{\lambda_{3}}-\sqrt{\lambda_{4}},\right\},
\end{equation}
where $\{\lambda_{i}\}$ are the eigenvalues of the product matrix $\rho \tilde{\rho}$ in decreasing order. For the simple form of the reduced density matrix given in Eq. (5), concurrence reduces to
\begin{equation}
 C=2\max\{0, |\rho_{14}|-|\rho_{22}|, |\rho_{23}|-\sqrt{\rho_{11}\rho_{44}}\}.
\end{equation}

\section{Correlations in the XY Model}

We start this section with the analysis of the thermal quantum and total correlations in the one-dimensional XY spin chain in transverse magnetic field. The Hamiltonian of the model is given by
\begin{equation}
 H_{XY}=-\frac{\lambda}{2}\sum_{j=1}^{N} [(1+\gamma)\sigma_j^x \sigma_{j+1}^x +(1-\gamma)\sigma_{j}^{y}\sigma_{j+1}^{y}]-\sum_{j=1}^{N}\sigma_{j}^{z}
\end{equation}
where $N$ is the number of spins, $\sigma_j^{\alpha}$ ($\alpha =x,y,z$) is the usual Pauli operators for a spin-$1/2$ at $j$th site, $\gamma$ ($0\leq\gamma\leq 1$) is the anisotropy parameter and $\lambda$ is the strength of the inverse external field. For $\gamma =0$ the above Hamiltonian corresponds to the XX model. When $\gamma\geq 0$ it is in the Ising universality class, and reduces to the Ising Hamiltonian in a transverse field for $\gamma=1$. We are interested in the region where the XY model exhibits two phases, a ferromagnetic and a paramagnetic phase, which are separated by a second-order QPT at the CP
$\lambda_c =1$. In the thermodynamic limit, the XY model can be solved exactly via a Jordan-Wigner map followed by a Bogoluibov transformation. Reduced density matrix of two spins $i$ and $j$ depends only on the distance between them, $r=|i-j|$, due to the translational invariance of the system.
\begin{figure}[H]
\begin{center}
\includegraphics[scale=0.56]{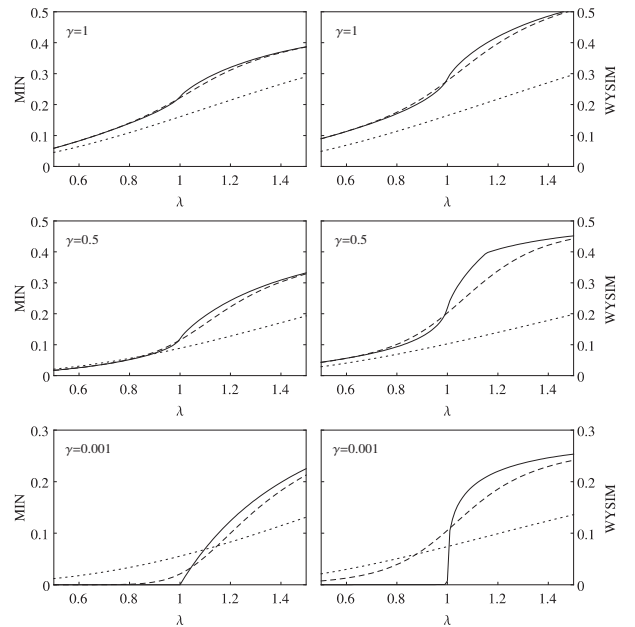}
\caption{The thermal total correlations quantified by MIN and WYSIM as a function of $\lambda$ for $\gamma=0.001,0.5,1$ at $kT=0$ (solid line), $kT=0.1$ (dashed line) and $kT=0.5$ (dotted line). The graphs are for first nearest neighbors.}
\end{center}
\end{figure}
\noindent The Hamiltonian is also invariant under parity transformation, meaning it exhibits $Z_2$ symmetry. Taking these properties into account, and neglecting the effects of spontaneous symmetry breaking (which are studied in Ref. [30-34]), the two-spin reduced density matrix of the system is given by [13]
\begin{equation}
 \rho_{0,r}=\frac{1}{4}[I_{0,r}+\langle\sigma^z\rangle (\sigma_0^z+\sigma_r^z)]+\frac{1}{4}\sum_{\alpha =x,y,z}\langle\sigma_0^{\alpha}\sigma_r^{\alpha}\rangle
 \sigma_0^{\alpha}\sigma_r^{\alpha},
\end{equation}
where $I_{0,r}$ is the four-dimensional identity matrix. The transverse magnetization is given by [41]
\begin{equation}
 \langle\sigma^z\rangle =-\int_0^{\pi} \frac{(1+\lambda\cos \phi)\tanh(\beta\omega_{\phi})}{2\pi\omega_{\phi}}d\phi,
\end{equation}
where $\omega_{\phi}=\sqrt{(\gamma\lambda\sin\phi)^2+(1+\lambda\cos \phi)^2}/2$, $\beta =1/k_bT$ with $k_b$ being the Boltzmann constant and T is the absolute temperature. Two-point correlation functions are defined as [42]
\begin{equation}
 \langle\sigma_0^x\sigma_r^x\rangle = \begin{vmatrix}
 G_{-1} & G_{-2} & \cdots & G_{-r} \\
 G_0 & G_{-1} & \cdots & G_{-r+1} \\
 \vdots & \vdots & \ddots & \vdots \\
 G_{r-2} & G_{r-3} & \cdots & G_{-1} \end{vmatrix},
\end{equation}
\begin{equation}
 \langle\sigma_0^y\sigma_r^y\rangle = \begin{vmatrix}
 G_1 & G_0 & \cdots & G_{-r+2} \\
 G_2 & G_1 & \cdots & G_{-r+3} \\
 \vdots & \vdots & \ddots & \vdots \\
 G_r & G_{r-1} & \cdots & G_1 \end{vmatrix},
\end{equation}
\begin{equation}
 \langle\sigma_0^z\sigma_r^z\rangle =\langle\sigma^z\rangle ^2-G_rG_{-r},
\end{equation}
where
\begin{align}
 G_r= &\int_0^{\pi} \frac{\tanh(\beta\omega_{\phi})\cos(r\phi)(1+\lambda\cos \phi)}{2\pi\omega_{\phi}}d\phi \\ \nonumber
      &-\gamma\lambda\int_0^{\pi} \frac{\tanh(\beta\omega_{\phi})\sin(r\phi)\sin(\phi)}{2\pi\omega_{\phi}}d\phi .
\end{align}
In Fig. 1, we present our results regarding the thermal total correlations quantified by MIN and WYSIM for first nearest neighbors as a function of $\lambda$ for $kT=0, 0.1, 0.5$ and $\gamma=0.001, 0.5, 1$. We note that although MIN and WYSIM behave in a similar fashion for $\gamma=1$, they show qualitatively different behaviors in the case of $\gamma=0.001$. Namely, WYSIM experiences a more dramatic increase about the CP $\lambda=1$ than MIN, and reaches to a constant value more quickly. Furthermore, it is also important to observe that as temperature increases, both of the measures cease to exhibit a non-trivial behavior about the CP.

It has been shown that QPTs can be characterized by looking at the two-spin reduced density matrix and its derivatives with respect to the tuning parameter driving the transition [14,15]. Since correlation measures are directly determined from the reduced density matrix, they provide information about the CPs and the order of QPTs. The CP for a second-order QPT at zero temperature is signalled by a divergence or discontinuity in the first derivative of the correlation measures. If the first derivative is discontinuous, then the divergence of the second derivative pinpoints the CP [14-16]. In Fig. 2, we plot the derivatives of MIN and WYSIM as a function of $\lambda$ for $kT=0, 0.1, 0.5$ and $\gamma=0.001, 0.5, 1$. We observe that both of the measures are capable of spotlighting the CP at $kT=0$ for all values of $\gamma$. It is worth to note that with increasing temperature, the divergence at CP disappears and the peaks of the derivatives start to shift. Therefore, the measures lose their significance in determining the CP of the transition.

We now turn our attention to the analysis of thermal quantum correlations quantified by OMQC and concurrence. In Fig. 3 and Fig. 4, we plot these measures and their derivatives with respect to the driving parameter $\lambda$ for first nearest neighbors as a function of $\lambda$ for $kT=0, 0.1, 0.5$. While concurrence suffers a drastic decrease as temperature increases, OMQC still captures significant amount of correlation, making it more robust against thermal effects. It can also be seen that at $kT=0$ the CP can be detected by analyzing the non-analyticities in the first derivatives of the measures.

Next, we discuss the question of whether the studied correlation measures can signal the emergence of  non-trivial product ground state in the XY spin chain. Despite the fact that the ground state of the model is entangled in general, for some special values of $\gamma$ and $\lambda$, the ground state becomes completely factorized. In particular, except the trivial factorization points $\lambda=0$ and $\lambda\rightarrow\infty$, there also exists a non-trivial factorization line corresponding
\begin{figure}[H]
\begin{center}
\includegraphics[scale=0.56]{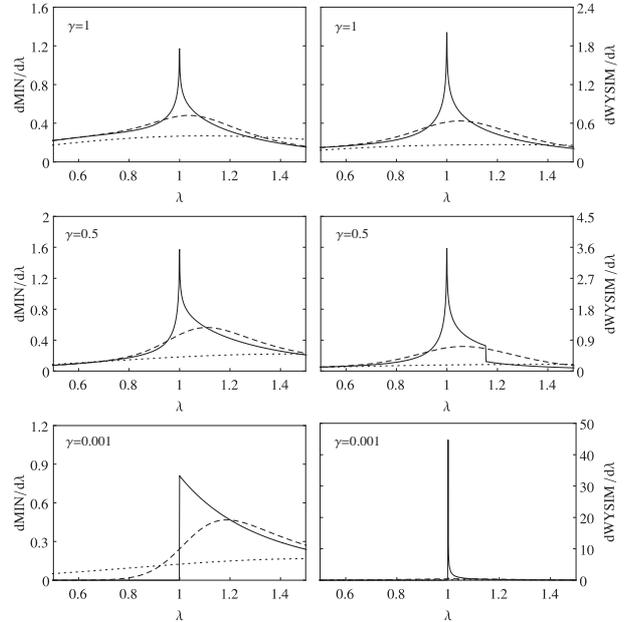}
\caption{The first derivatives of MIN and WYSIM as a function of $\lambda$ for $\gamma=0.001,0.5,1$ at $kT=0$ (solid line), $kT=0.1$ (dashed line) and $kT=0.5$ (dotted line). The graphs are for first nearest neighbors.}
\end{center}
\end{figure}
\noindent  to $\gamma^2+\lambda^{-2}=1$. Accordingly, as can seen from the behavior of concurrence in Fig. 3 for $\gamma=0.5$, entanglement vanishes at $\lambda\simeq1.15$, which spotlights the occurrence of a product ground state. It is shown in Fig. 2 that, unlike OMQC and MIN, WYSIM can signal this factorization point through a non-analytical behavior in its derivative. For QD to identify this point when the distance between the spins is fixed, the effects of SSB must be taken into account [33,34,43]. Therefore, it is important to recognize that the calculation of WYSIM between the spins at a fixed distance enables us to detect the product ground state even in the absence of SSB.

Having discussed the behaviors of the thermal total and quantum correlations, we now explore the ability of these measures to correctly estimate the CP of the QPT at finite temperature. Despite the disappearance of the singular behavior of MIN, WYSIM, OMQC and concurrence with increasing temperature, it might still be possible to estimate the CP at finite temperature [28]. For sufficiently low temperatures, divergent behaviors of the first derivatives of correlation measures at $T=0$ will be replaced by a local maximum or minimum about the CP. Therefore, in order to estimate the CP, we search for this extremum point. On the other hand, a discontinuous first derivative at $T=0$ requires us to look for an extremum point in the second derivative for $T>0$. In Fig. 5, we present the results of our analysis regarding the estimation of CP as a function of $kT$ for first and second nearest neighbors when $\gamma=0.001,0.5,1$. Before starting to compare the ability of MIN, WYSIM, OMQC and concurrence to indicate the CP in detail, we notice that the success rates of these
\begin{figure}[H]
\begin{center}
\includegraphics[scale=0.56]{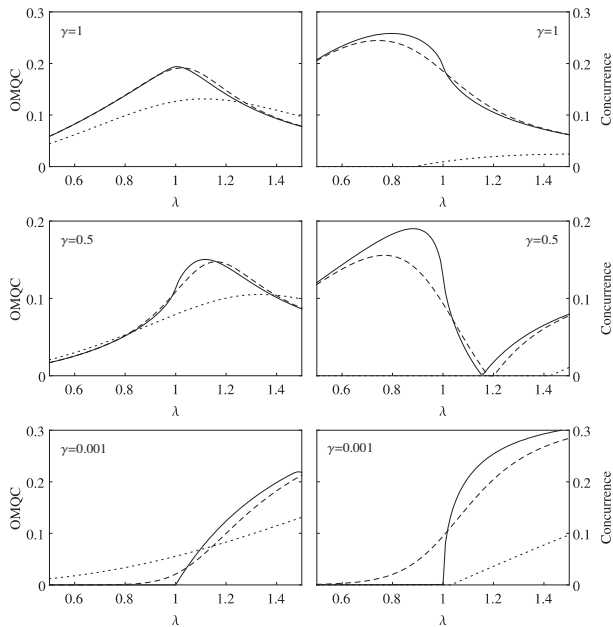}
\caption{The thermal quantum correlations quantified by OMQC and concurrence as a function of $\lambda$ for $\gamma=0.001,0.5,1$ at $kT=0$ (solid line), $kT=0.1$ (dashed line) and $kT=0.5$ (dotted line). The graphs are for first nearest neighbors.}
\end{center}
\end{figure}
\begin{figure}[H]
\begin{center}
\includegraphics[scale=0.56]{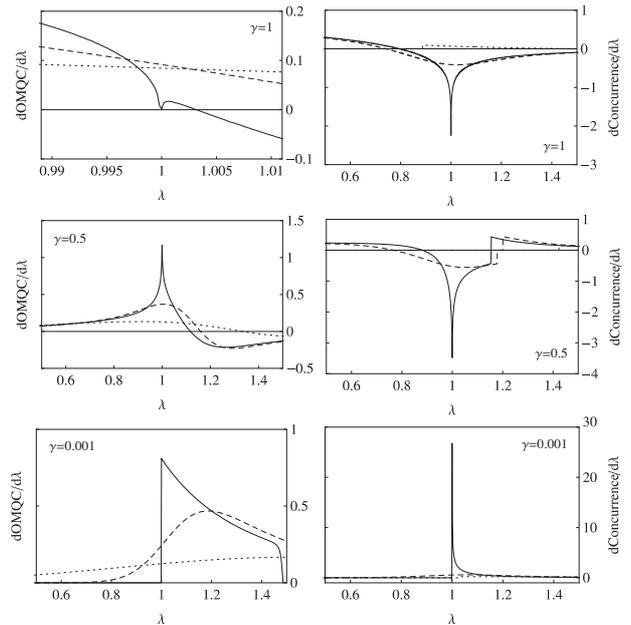}
\caption{The first derivatives of OMQC and concurrence as a function of $\lambda$ for $\gamma=0.001,0.5,1$ at $kT=0$ (solid line), $kT=0.1$ (dashed line) and $kT=0.5$ (dotted line). The graphs are for first nearest neighbors.}
\end{center}
\end{figure}
\noindent measures strongly depend on the anisotropy parameter of the Hamiltonian.

In the case of first nearest neighbors, at $\gamma=1$, all of the correlation measures are able to predict the CP reliably, with concurrence being the most effective among them. When $\gamma=0.5$ MIN turns out to be the worst CP estimator. While WYSIM and concurrence points out the CP relatively well as compared to MIN, OMQC clearly outperforms all others and estimates the CP in a exceptionally accurate way. For $\gamma=0.001$, MIN and OMQC become identical, and they predict the location of the CP significantly worse than WYSIM and concurrence.

For second nearest neighbors, even though we do not present the graphs of correlation measures and their derivatives, the CP has been inspected by performing the same analysis as in the first nearest neighbor case. The CPs estimated by WYSIM, OMQC and MIN for $\gamma=1$ deviate from the true CP by the same amount but they are still acceptable. In the case of $\gamma=0.5$, both concurrence and OMQC estimate the CP very well in contrast to WYSIM and MIN. Finally, when $\gamma=0.001$, while WYSIM and concurrence spotlight the CP remarkably well, OMQC and MIN perform very poorly. It is also worth to notice that concurrence performs even better than the first nearest neighbors case for $\gamma=0.5$ and $\gamma=0.001$.

Furthermore, the ability of entanglement of formation (EOF) and QD to estimate the CP of the XY spin chain at finite temperature has been recently studied by Werlang et al. [28]. The performance of the correlations measures used in this work as compared to QD and EOF depend on the anisotropy parameter of the Hamiltonian and also on the distance between the spin pairs. For instance, in the first nearest neighbors case at $\gamma=0.5$, among the correlation measures considered here, only OMQC performs as well as QD and EOF. On the other hand, for the second nearest neighbors at $\gamma=0.001$, while WYSIM and concurrence turn out to be better CP estimators than QD and EOF, MIN and OMQC do not perform as well. We lastly note that apart from a limited number of special cases, QD still proves to be the most accurate CP estimator for the anisotropic XY spin chain.
\begin{figure*}[ht!]
\begin{center}
\includegraphics[scale=0.76]{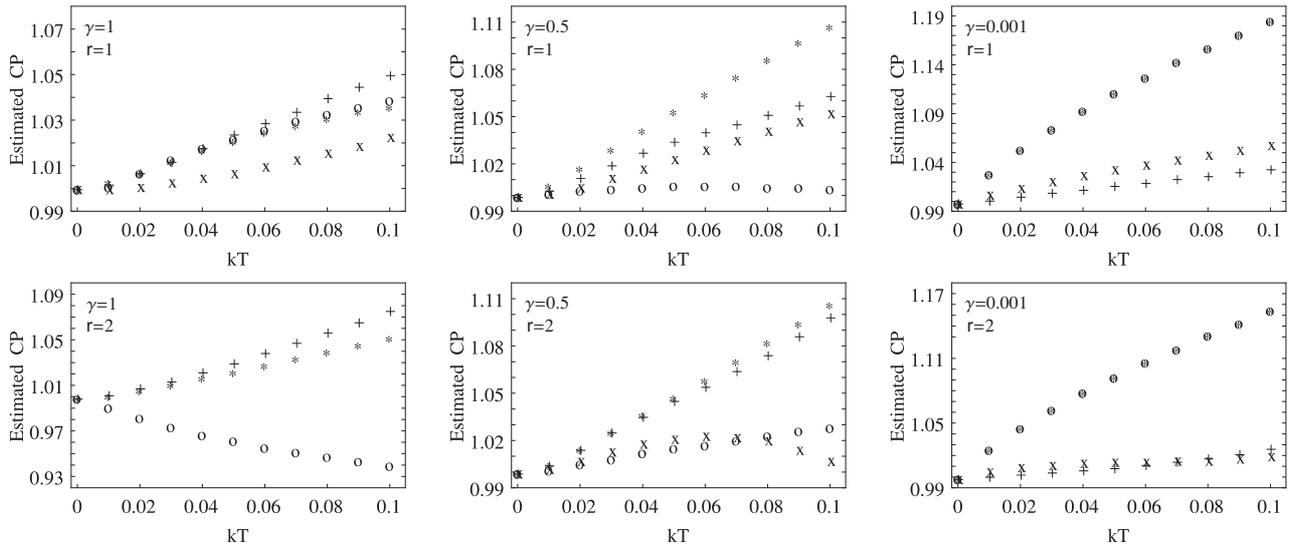}
\caption{The estimated values of the CP as a function of $kT$ for three different values of the anisotropy parameter $\gamma=0.001,0.5,1$. The CPs in the graphs are estimated by OMQC (denoted by o), WYSIM (denoted by +), MIN (denoted by $\ast$) and concurrence (denoted by x). Concurrence is not included for $\gamma=1$ and $r=2$, since it vanishes at even very low temperatures.}
\end{center}
\end{figure*}

Inspired by the methods of Ref. [21], we now examine the long-range behavior of the thermal total and quantum correlations for the one-dimensional XY model in transverse magnetic field. While entanglement vanishes for distant spin pairs even in the ordered ferromagnetic phase, QD has been shown to remain non-zero [18, 21]. Fig. 6 demonstrates our results related to the dependence of MIN, WYSIM and OMQC on the distance between the spin pairs at finite temperature, for $\lambda=0.75,0.95,1.05,1.5$ and $\gamma=0.001,1$. In case of $\gamma=0.001$, neither of the correlation measures remain significant when the distance between the spin pairs is increased. We can also see that the decay of the correlations hasten when the temperature rises. For the Ising model limit ($\gamma=1$), even though MIN, WYSIM and OMQC approach to a finite value in the ordered phase for sufficiently low temperatures, thermal effects wipe out the correlations between distant spin pairs after a certain temperature.

\section{Conclusion}

In summary, we have discussed the thermal quantum and total correlations in the one-dimensional anisotropic XY model in transverse magnetic field from several perspectives. We have quantified the correlations using recently proposed correlation measures such as WYSIM, MIN and OMQC, and a well known entanglement measure concurrence. Analyzing these measures in the parameter space of the Hamiltonian for first and second nearest neighbors, we have found that all of the considered measures are capable of indicating the CP of the transition. Although the interesting behavior of the measures in the vicinity of the CP disappears as the temperature increases, for sufficiently low temperatures, it is still possible to estimate the CP by looking at the derivatives of the correlation measures. We have observed that the ability of the measures to predict the CP strongly depend on the anisotropy parameter $\gamma$. For instance, while OMQC spotlights the CP with a remarkably high accuracy at $\gamma=0.5$ for first nearest neighbors, it performs very poorly at $\gamma=0.001$. On the other hand, WYSIM points out the CP reasonably well at $\gamma=0.001$ for both first and second neighbors. Moreover, we have shown that, among the new measures considered in this work, only WYSIM is able to identify the factorization point of the XY spin chain even if we disregard the effects of SSB. These results demonstrate for the first time that OMQC and WYSIM are relevant quantities for identifying CPs in concrete physical problems. Next, we have investigated how WYSIM, MIN and OMQC are affected as we increase the distance between spin pairs. We have found that the case of $\gamma=0.001$ is more susceptible to both increasing distance of spin pairs and thermal effects. It would also be interesting to study these measures for different quantum critical spin systems such as XXZ chain. Finally, we leave the discussion of the effects of spontaneous symmetry breaking on these correlation measures as a future work.

\section*{Acknowledgements}
We would like to thank two anonymous referees for their valuable comments. This work has been partially supported by the Scientific and Technological Research Council of Turkey (TUBITAK) under Grant 111T232.

\begin{figure}[H]
\begin{center}
\includegraphics[scale=0.63]{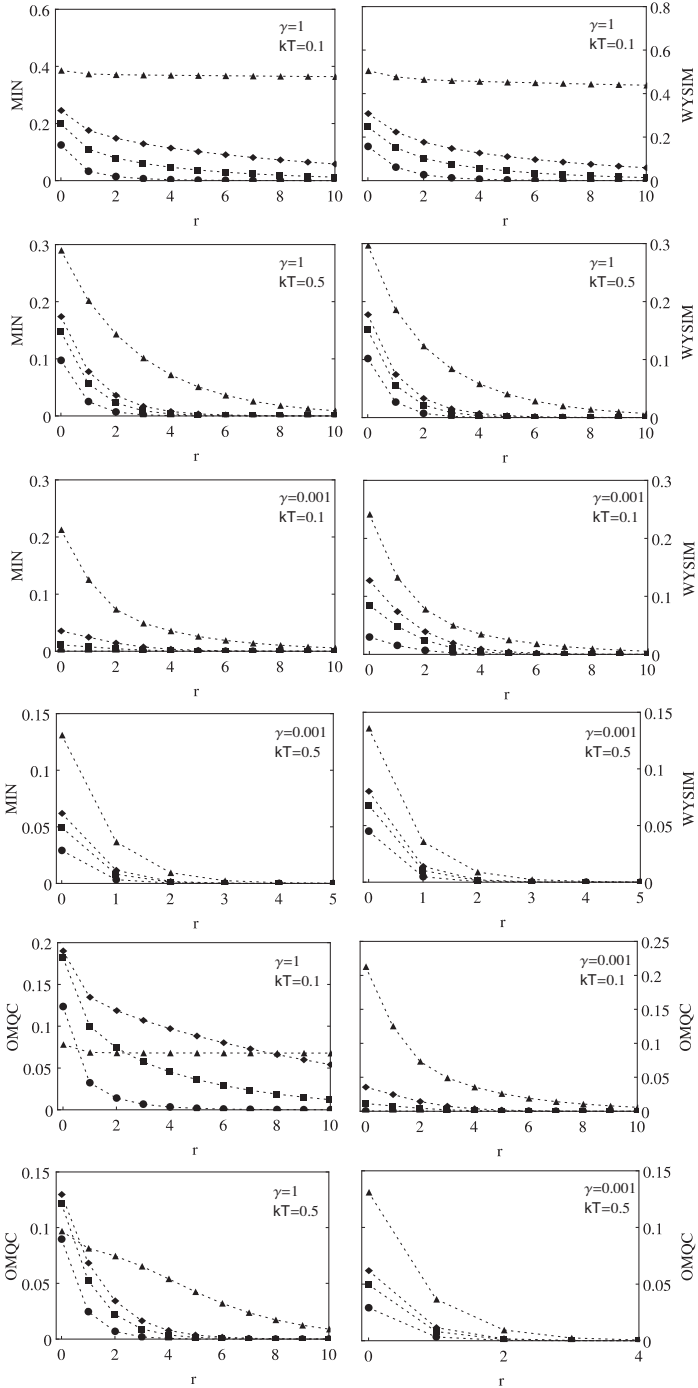}
\caption{Long-range behavior of the thermal total and quantum correlations for $\gamma=0.001$ and $\gamma=1$ at $kT=0.1, 0.5$. The circles, squares, diamonds and triangles correspond to $\lambda=0.75$, $\lambda=0.95$, $\lambda=1.05$ and $\lambda=1.5$, respectively. }
\end{center}
\end{figure}


\begin{thebibliography}{99}
 \bibitem{horodecki09} R. Horodecki, P. Horodecki, M. Horodecki, K. Horodecki, Rev. Mod. Phys. 81 (2009) 865.
 \bibitem{amico08} L. Amico, R. Fazio, A. Osterloh, V. Vedral, Rev. Mod. Phys. 80 (2008) 517.
 \bibitem{hendenson01} L. Hendenson, V. Vedral, J. Phys. A: Math. Gen. 34 (2001) 6899.
 \bibitem{ollivier02} H. Ollivier, W. H. Zurek, Phys. Rev. Lett. 88 (2002) 017901.
 \bibitem{datta08} A. Datta, A. Shaji, C. M. Caves, Phys. Rev. Lett. 100 (2008) 050502.
 \bibitem{knill98} E. Knill, R. Laflamme, Phys. Rev. Lett. 81 (1998) 5672.
 \bibitem{modi12} K. Modi, A. Brodutch, H. Cable, T. Paterek, V. Vedral arXiv:1112.6238v1.
 \bibitem{dakic10} B. Daki\'{c}, V. Vedral, \v{C}. Brukner, Phys. Rev Lett. 105 (2010) 190502.
 \bibitem{luo11} S. Luo, S. Fu, Phys. Rev. Lett. 106 (2011) 120401.
 \bibitem{luo12} S. Luo, S. Fu, C. H. Oh, Phys. Rev. A 85 (2012) 032117.
 \bibitem{girolami12} D. Girolami, G. Adesso, Phys. Rev. Lett. 108 (2012) 150403.
 \bibitem{sachdev} S. Sachdev, Quantum Phase Transitions, Cambridge University Press, Cambridge, 2011.
 \bibitem{osborne02} T. J. Osborne, M. A. Nielsen, Phys. Rev. A 66 (2002) 032110.
 \bibitem{dillenschneider08} R. Dillenschneider, Phys. Rev. B 78 (2008) 224413.
 \bibitem{wu04} L.-A. Wu, M. S. Sarandy, D. A. Lidar, Phys. Rev. Lett. 93 (2004) 250404.
 \bibitem{sarandy09} M. S. Sarandy, Phys. Rev. A 80 (2009) 022108.
 \bibitem{batle10} J. Batle, M. Casas, Phys. Rev. A 82 (2010) 061201.
 \bibitem{maziero10} J. Maziero, H. C. Guzman, L. C. C\'{e}leri, M. S. Sarandy, R. M. Serra, Phys. Rev. A 82 (2010) 012106.
 \bibitem{liu11} B.-Q. Liu, B. Shao, J.-G Li, J. Zou, L.-A. Wu, Phys. Rev. A, 83 (2011) 052112.
 \bibitem{li12} B. Li, Y.-S. Wang, Physica B, 407 (2012) 77.
 \bibitem{maziero12} J. Maziero, L. C. C\'{e}leri, R. M. Serra, M. S. Sarandy, Phys Lett. A 376 (2012) 1540.
 \bibitem{justino11} L. Justino, T. R. de Oliveira, Phys. Rev. A 85 (2012) 052128.
 \bibitem{cheng12} W. W. Cheng, C. J. Shan, Y. B. Sheng, L. Y. Gong, S. M. Zhao, B. Y. Zheng, Physica E 44 (2012) 1320.
 \bibitem{altintas12} F. Altintas, R. Eryigit, 	arXiv:1202.1495v2.
 \bibitem{rulli11} C. C. Rulli, M. S. Sarandy, Phys. Rev. A 84 (2011) 042109.
 \bibitem{li11} Y.-C. Li, H.-Q. Lin, Phys Rev. A 83 (2011) 052323.
 \bibitem{werlang10} T. Werlang, C. Trippe, G. A. P. Ribeiro, G. Rigolin, Phys. Rev. Lett. 105 (2010) 095702.
 \bibitem{werlang11} T. Werlang, G. A. P. Ribeiro, G. Rigolin, Phys. Rev A 83 (2011) 062334.
 \bibitem{kurmann82} J. Kurmann, H. Thomas, G. M\"{u}ller, Physica A 112 (1982) 235.
 \bibitem{oliveira08} T. R. de Oliveira, G. Rigolin, M.C. Oliveira, E. Miranda, Phys. Rev. A 77 (2008) 032325.
 \bibitem{syljuasen03} O. F. Sylju{\aa}sen, Phys. Rev. A 68 (2003) 060301(R).
 \bibitem{osterloh06} A. Osterloh, G. Palacios, S. Montangero, Phys. Rev. Lett. 97 (2006) 257201.
 \bibitem{tomasello11} B. Tomasello, D. Rossini, A. Hamma, L. Amico, Europhys. Lett. 96 (2011) 27002.
 \bibitem{saguia11} A. Saguia, C. C. Rulli, T. R. de Oliveira, M. S. Sarandy, Phys. Rev. A 84 (2011) 042123.

 \bibitem{hill97} S. Hill, W. K. Wootters, Phys. Rev. Lett. 78 (1997) 5022.
 \bibitem{wootters98} W. K. Wootters, Phys. Rev. Lett. 80 (1998) 2245.
 \bibitem{wigner63} E. P.Wigner and M. M. Yanase, Proc. Natl. Acad. Sci. USA 49 (1963) 910.
 \bibitem{cui10} J. Cui, J.-P. Cao, H. Fen, Phys. Rev. A 82 (2010) 022319.
 \bibitem{allegra11} M. Allegra, P. Giorda, A. Montorsi, Phys. Rev. B 84 (2011) 245133.
 \bibitem{anfossi05} A. Anfossi, P. Giordia, A. Montorsi, F. Traverse, Phys. Rev. Lett. 95 (2005) 056402.
 \bibitem{barouch70} E. Barouch, B. M. McCoy, M. Dresden, Phys. Rev A 2 (1970) 1075.
 \bibitem{barouch71} E. Barouch, B. M. McCoy, Phys. Rev A 3 (1971) 786.
 \bibitem{sarandy12} M. S. Sarandy, T. R. de Oliveira, L. Amico, arXiv:1208.4817v1.
 \end{thebibliography}
\end{document}